# Nucleation and growth of catalyst-free ZnO nano-structures

Jai Singh, Anchal Srivastava, R.S. Tiwari and O. N. Srivastava*

Department of Physics, Banaras Hindu University, Varanasi, India-221005

**Abstract**

This paper deals with the investigations on the nucleation and growth of ZnO nanostructures in a catalyst free synthesis. The ZnO nanostructures have been formed by evaporation of Zn (99.99%) in $O_2$ and Ar atmosphere in single zone furnace under two temperature regions, region A (~1173-1073K) and region B (~873-773K). Through application of XRD and TEM techniques, it has been shown that first ZnO is formed which changes to $ZnO_x$ through creation of oxygen vacancies. The $ZnO_x$ acts as self-catalyst and leads to formation of various nanostructures. Those observed in the present investigation are nanotetrapods (1D, diameter~70-450nm, length~2-4.5μm) nanorods (1D, diameter~45-95nm, length~2.5-4.5μm), nanoflowers(2D,central core diameter~90-185nm, length of petals/nanorod~1.0-3.5μm) and nanoparticles (3D, size~ 0.85-2.5μm). These nanostructures have been revealed by SEM explorations. Attempts have been made to explain the formation of the various nanostructures in terms of the creation and distribution of the $ZnO_x$, the temperature as well as oxygenation conditions.

**Keywords:** Self catalyst, ZnO nanostructures, Lattice parameters

Corresponding author

Fax: +91-542-2368468; email: hepons@yahoo.com

# 1. Introduction

Out of several nanomaterials studied so far carbon nanotubes [1,2] and ZnO [3,4] exhibit the widest varieties of nanostructures. ZnO is presently hotly pursued material in regard to formation of myriad nanostructures. Some of these are nanorods, nanobelts, nanorings, nanocombs, nanostars, nanonails etc. [3-6] It may be mentioned that extended and oriented nanostructures are of potential use for many applications such as microelectronic devices, light-emitting diodes, drug delivery, optical storage, chemical and biological sensors, diagnostic markers, energy conversion and storage etc. [3,5,6,7]

Various synthesis methods capable of formation of different types of ZnO nanostructures have been demonstrated [3-8]. The central aims of the present investigations are to find out the reason for the nucleation and growth of ZnO nanostructures without any catalytic agents. This aspect does not seem to have been properly addressed so far. We have synthesized ZnO nanostructures through evaporation of Zn powder under $O_2$ flow rate (800 sccm; standard cubic centimeters per minute) and Ar flow rate (500 sccm). We have established two temperature regions for the silica growth tube in a furnace. One of these termed as high temperature (1173-1073K) region is near the alumina boat containing Zn powder. This has been termed as region "A" and the other, the low temperature region (873-773K) as "B". We have shown that the ZnO deposits first as stoichiometric ZnO and later it transforms to $ZnO_x$ through creation of oxygen vacancies. The formation of $ZnO_x$ has been confirmed through extensive exploration employing XRD and the TEM techniques. The various nanostructures namely the nanoparticles, nanotetrapod networks, nanorods and nanoflowers are believed to nucleate on the $ZnO_x$ nuclei, which work as self catalyst. The relationship between nucleation, growth and nanomorphologies are helpful for controlling the physical properties of nanostructures. Keeping this in view, attempts have been made to correlate the morphology and the details of $ZnO_x$ including the distribution of oxygen vacancies forming the $ZnO_x$.

**2. Experimental, Results and Discussions**

It may be mentioned that nucleation and formation of ZnO nanostructures have been proposed to take place through two types of catalysts. In one of these foreign impurities are added to ZnO which act as catalysts and nucleate the nanostructures. Thus elemental particles e.g. Au, Sn, C, or some additives such as $In_2O_3$, NiO, $Ga_2O_3$ etc. are taken to act as catalyst particles, which help in nucleation of ZnO nanostructures.[3-5, 10] The second catalysts are Zn,$ZnO_x$ either independently or in conjunction which are thought to be responsible for the nucleation and growth of ZnO nanostructures.[7,11]

In the present case pure Zn (Thomas Baker 99.99% purity) was employed for evaporation under $O_2$ and Ar gases to form ZnO nanostructures. Therefore, the presence of foreign impurity particles to act as catalyst can be ruled out. In regard to the known self catalysts Zn or $ZnO_x$, we have carried out thorough search. The ZnO, which was evaporated under $O_2$ flow rate (800 sccm), Ar flow rate (500 sccm) ambient was deposited in a silica tube (diameter 3cm, length 60cm), which was put in a single zone furnace. The temperature of the furnace was adjusted so as to obtain two temperature regions. These regions are A with temperature (1173-1073K) and region B (873-773K). The ZnO material from regions A and B were collected and subjected to structural characterization.

The resulting ZnO material was analyzed through XRD, SEM and TEM. Fig. 1(a) and Fig. 1(b) show representative XRD patterns of the ZnO obtained through evaporation of Zn powder. Whereas Fig. 1(a) shows ZnO obtained from region A, Fig. 1(b) corresponds to the ZnO collected from region B. Detailed analysis of these patterns did not show any presence of Zn. This was found to be true for ZnO obtained from both the regions A and B. Extensive search for Zn nuclei or droplets employing SEM and TEM also did not succeed. Next, we set out to explore the presence of $ZnO_x$. It is expected that the presence of oxygen vacancies leading to $ZnO_x$ will result in variation of lattice parameters 'a' and 'c' of the as deposited material as compared to the parameters of the standard ZnO. The XRD patterns of the samples were evaluated employing PW-1710 (CuKα, 30kV, 20mA). The K$\alpha_1$ and K$\alpha_2$ peaks were separated through a graphite monochormator. The calibration of the diffractometer was done by using Si powder ($d_{111}$=3.1353Å). The accurate lattice parameters were obtained employing least square fitting based computer programme. Analysis of XRD patterns Fig. 1(a) and Fig.1 (b) suggested that the lattice parameter for the ZnO obtained from region A and B are different. These are a=3.251Å and c=5.208Å for region A (1173-1073K) and a=3.265Å and c=5.229Å for region B. These lattice parameters are listed in Table 1. Both these lattice parameters are different than the standard JCPDS parameters for ZnO, a=3.2498Å and c=5.1948Å.[12, 13,14] This is suggestive of formation of nonstoichiometric $ZnO_x$.[13] Inset to Fig. 2 shows the overlapped region of Fig. 1(a) and Fig. 1(b) for 2θ ≈30° to ≈60°. This clearly shows the variation of lattice parameters as brought out by shift in corresponding peaks.

Repeated experiments spread on several runs invariably showed different lattice parameters for regions A and B and their estimates were very close to those outlined above. As is known, the presence of oxygen deficiency leads to variation of lattice parameters.[13] Also ZnO being ionic crystal the lattice parameters increase with oxygen deficiency.[13, 17] Table 1 clearly shows that the lattice parameters for both regions A (1173-1073K) and region B (873-773K) are higher than the standard JCPDS (36-1451) values. Also the increase in 'c' parameter is about 20 times higher than 'a' parameter for region A and by about 6 times for region B. As can be seen, the variation in 'c' parameter of ZnO from the standard $ZnO_x$ parameter is significantly larger than the variation of 'a' parameter. This is, however, expected because of the creation of oxygen vacancies.[13, 16,18] It may be pointed out that variation in lattice parameter for ZnO prepared under different conditions have also been reported earlier. However, this variation was taken to reflect the state of strain of ZnO nanostructures.[18]

There could be two situations for the formation of $ZnO_x$ particles. In one of these $ZnO_x$ is formed due to partial oxidation of Zn in the stream of $O_2$ and then get deposited on the walls of growth tube. In the other, full oxidation of Zn in the $O_2$ stream leads to formation of ZnO. This stoichiometric ZnO then looses oxygen to lead to $ZnO_x$ through the reaction

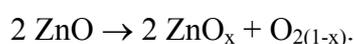
$$2\, ZnO \rightarrow 2\, ZnO_x + O_{2(1-x)}.$$

In order to find out which of these possibilities is really operative in the present case, the Zn was evaporated for a very short time span of ~2minutes(min.) instead of ~10min. The lattice parameter of ZnO deposited for 2 min. was evaluated. Investigations spread on several samples showed that invariably the lattice parameter of the 2 min. deposited ZnO sample was very close to JCPDS (36-1451) value. Representative values of 'a' and 'c' parameter for ZnO deposited for 2 min. is shown in Table 1. These observations suggest that in the initial stages nearly stoichiometric ZnO is formed. This ZnO then later develops oxygen deficiency to form $ZnO_x$ nuclei. In passing it may be mentioned that the presence of only Zn was not observed for any of the deposition runs.

In order to further confirm the formation of zinc suboxide $ZnO_x$, we explored ZnO from both regions A and B employing selected area electron diffraction (SAED). These investigations revealed two different types of SAED. In passing it should be pointed out that the ZnO lattice parameters 'a' and 'c' obtained through selected area electron diffraction patterns from region A and B, were larger than the standard 'a' and 'c' parameters. The estimates for these were broadly compatible with those from the XRD data (Table 1). It must, however, be pointed out that the lattice parameters as obtained through electron diffraction are less accurate than those from XRD. One of prominent diffraction pattern obtained corresponds to (10.0) type diffraction pattern showing rather diffuse (00.1) type spots (Fig. 3). As shown earlier ZnO is converted due to creation of oxygen vacancies. These vacancies when in disordered state will produce diffuse spots as has been actually observed.[17, 19] The SAED patterns obtained from both regions A and B exhibited diffuse (00.l) spots. It should be pointed out that the diffuseness of the (00.1) type spots (some of these are marked by arrows) are not due to EM lense settings etc. Since these were already optimized before taking the SAED. Yet another type of diffraction pattern, which provides further evidence for the existence of ZnOx phase, was the SAED showing the presence of modulated structures. Representative example of such diffraction patterns is shown in Fig. 4. The modulated[25] phases were found from ZnO collected from both regions A and B. As is indicated in the figure, there is a modulation of period brought out by the presence of three equally spaced additional spots along the [00.1] direction. Thus the 'c' parameter of the modulated phase is four times as compared to the parent phase. Since as outlined earlier ZnO after its formation develops oxygen vacancies, the ordering of oxygen vacancies will result in the formation of modulated phase. In view of the observation of the vacancy ordered $ZnO_x$ phases, it can be taken that the diffuse spots in Fig. 3 reveal, the disordered oxygen vacancy phase of ZnO ($ZnO_x$).

Having established the existence of $ZnO_x$, which as per known results, would act as self catalyst, [7,11] we proceed now to look into the formation of ZnO nanostructures. As described earlier, the as deposited ZnO material was collected from two regions of the growth tube namely the region A (1173-1073K) and the region B (873-773K). In order to discern the initial stages of nanostructure formation, we carried out initially the deposition only for few minutes (~2 min.). The rest of the depositions were done for longer durations (~ 10 min.), so that the deposition, transformation and the nanostructure formation gets completed. Investigations of the morphologies of the as synthesized deposited ZnO nanostructure were carried out through the techniques of scanning electron microscopy employing secondary electron imaging.

The morphologies of the initial stages (deposition for ~2 min.) for high temperature region A and B consists of ZnO islands. These islands had three dimensional characteristics. A representative example of islands structure is brought out by Fig. 5. Some of the clearly discernible three dimensional islands have been marked by arrows. It can also be noticed that the islands maintain their separate entity and are not interconnected. The ZnO deposits collected after ~10 minutes growth run, did not show any island structure. Instead they exhibited several variety of nanostructures. It can, therefore, be taken that the islands after developing oxygen deficiency and formation of $ZnO_x$ catalytic centers, yield to give rise to various nanostructure variants. These will now be described and discussed.

The dominant ZnO nanostructures formed under the higher temperature region A is in the form of tetrapods. Elucidative examples of this are shown by Figs. 6 and 7. As is easily discernible from these figures extensive interconnected network of tetrapods are present. One interesting aspect of tetrapod networks as evidenced by Figs. 6 and 7, which does not seem to have been noticed or described so far by other workers, is the presence of several layers tetrapod networks. These layers lie one on the top of the other. This feature is brought out by the presence of two different types of contrasts. We will, for making the formation of tetrapod networks layers intelligible, first consider the case of only two layered tetrapod network configuration. One corresponds to grey (dim) contrast of tetrapod networks, the other by white (bright) contrast of the overlaid tetrapod networks. This is present for tetrapod networks shown in both the Figs. 6 and 7. For clarity some of the tetrapod networks regions in the layer below is marked by 'L' and some portions of the overlaid layer by 'U' in the inset.

As regards the formation of tetrapod in keeping with the by now generally accepted mechanism of formation, of the four legs of tetrapods comprise of nanorods which grow from inversion type octotwin embryo. [20,21] The tetrapod network apparently is formed in the high temperature region (1173-1073K) where the formation of ZnO is fastest. We believe that in the initial stage of formation, the ZnO embryo is nucleated. This is then transformed to $ZnO_x$ through loss of oxygen. The $ZnO_x$ working as catalyst leads to nucleation followed by growth of the tetrapod nanostructure. High rate of oxidation of Zn leading to formation of ZnO which will then produce copious $ZnO_x$ nuclei. These $ZnO_x$ nuclei will acts as growth centers. The temperature and the oxygen flow is highest for this region. These are the conditions under which tetrapod structures are formed. [20-24] Tetrapod is the configuration embodying four legs which are formed by nanorods growing along the four easy direction of the type [00.1]. Once a layer of tetrapod is formed, because of high temperature (~1173-1073K), which still prevails for this region, the extreme tips of the tetrapod which is thinnest region develops nonstoichiometry through loss of oxygen. This results in the formation of $ZnO_x$. Which forms the nucleus for another tetrapod. Incoming ZnO gaseous fluxes will deposit on this $ZnO_x$ to form another tetrapod nanostructure.

Other tetrapods of the reference layer will similarly lead to tetrapod formation of another tetrapod network. This network of tetrapods will grow on top of the reference layer, which works like a substrate and forms the second layer of tetrapods. If the above described conditions of tetrapod formation still prevail, yet another tetrapod network layer similar to the second layer will form. When looked under scanning electron microscope, the top tetrapod network will produce white (bright) contrast whereas the network layer below this will produce grey (dim), and the layer still below grey (dimmer) contrast. This is in keeping with the experimental observations as shown by Figs. 6 and 7 where several layers of tetrapod networks one above the other can be easily seen to be present.

Figs 8 and 9 bring out the nanostructures grown in the temperature region B, the dominant nanostructures consists of a central nearly circular region on the periphery of which ZnO nanorods grow. This is in keeping with the known results on nanostructure formation of ZnO. The nanorods are the simplest ZnO nansotructure, which form due to rapid growth along one of the fastest growth direction i.e. [00.1]. The SAED of the region containing nanorods confirmed the orientation of nanorods as along [00.1] type directions. It may, however, be noticed that the ZnO nanorods grown in Figs. 8 and 9 have significant morphological differences in their configurations. Whereas Fig. 8 shows rather random distribution of nanorods, Fig. 9 brings out ordered array of nanorods aligned along two different [00.1] directions, which are at 90º apart. These are shown by NR and NR′ in Fig.9. As regards the region C in both the Figs. 8 and 9, here again the dominant nanostructures are nanorods but these are oriented nearly along the electron beam directions and hence instead of full lateral view of the nanorods only near end or inclined view is visible In addition to nanorods some particle like morphologies are also seen in Figs. 8 and 9.

In Fig. 8 the nanorod formation is all along the periphery of the central region, which is nearly circular. The nanorods are not aligned upto any perceptible coherence lengths. As described earlier the oxygen vacancies leading to the formation of $ZnO_x$ are responsible for growth of nanostructures. The oxygen vacancies may be in solid solution and hence disordered. It will apparently lead to formation of disordered $ZnO_x$ catalytic nuclei. The nanorods, which will grow on these disordered $ZnO_x$ nuclei, will naturally be randomly oriented. The formation of randomly oriented nanorods is shown in Fig. 8. The central region, which is nearly circular, also shows nanorod formation. However, the nanorods here are nearly along the electron beam i.e. are being seen nearly end on. Thus the nanorods produce only sporty contrast in the center. The said feature surrounded by nanorods distributed along the circumference of the central region produces a flower like figure. The total configuration in Fig.8 can be termed as flower nanostructure. The oxygen vacancies, however, can precipitate and get ordered producing aligned $ZnO_x$ catalytic nuclei. It may be mentioned that as already described, evidence for the existence of both ordered and disordered oxygen vacancies have been obtained through selected area electron diffraction studies. The presence of modulated structures formed by the presence of ordered oxygen vacancies (Fig.4), is expected to lead to growth of ordered / aligned nanorods. We believe this is the way through which the aligned rows of nanorods in Fig. 9 (e.g. at NR and NR′) get formed.

**Conclusions**

To conclude, it can be said that in the present investigation, we have shown that for the catalyst free formation of ZnO nanostructures, $ZnO_x$ acts as self catalyst. Initially ZnO is formed which transforms to $ZnO_x$ through loss of oxygen leading to the creation of oxygen vacancies. The evidence for the formation of $ZnO_x$ has been found through variation of lattice parameters as obtained from XRD patterns. The oxygen vacancies can be either disordered or ordered as evidenced through selected area electron diffraction patterns. The $ZnO_x$ nuclei lead to the growth of several ZnO nanostructures. These are nanoparticles, nanotetrapod networks, aligned and random distribution of nano-rods and the flower like nano configuration. Feasible reasons for the formation of the various ZnO nanostructures have been advanced in terms of formation of $ZnO_x$ and their distributions, as well as temperature and oxidation conditions.

**Acknowledgements**

The authors are grateful to Professor C.N.R. Rao, FRS, Professor S.K. Joshi, Professor A.R. Verma, Professor P. Ramachandra Rao, Professor A.K. Raychaudhari and Professor S. Lele for encouragement. The financial assistance from CSIR (Jai Singh) and DST (NSTI) are gratefully acknowledged.

**Table 1**

| Temperature (1173-1073K) | Temperature (873-773K) | Deposited Time (min) | a(Å) | c(Å) |
|---|---|---|---|---|
| Region A | | 10 | 3.2512±0.0008 | 5.2078±0.0018 |
| | Region B | 10 | 3.2651±0.0017 | 5.2292±0.0039 |
| Region B | | 2 | 3.2430±0.0020 | 5.2015±0.0045 |
| | Region B | 2 | 3.2487±0.0012 | 5.2023±0.0028 |

**Figure Captions**

Fig. 1: XRD pattern of ZnO from 'A' the high temperature region and 'B' low temperature region.

Fig. 2: Superposition of XRD pattern from regions A and B between $2\theta \approx 30°$ to $60°$. The presence of two sets of closely spaced XRD peaks reveal the difference in the lattice parameter of ZnO from regions 'A' and 'B'.

Fig. 3: (10.0) type diffraction pattern. Notice the presence of (00.1) type diffuse diffraction spots.

Fig. 4: SAED pattern of ZnO modulated phase bringing out the presence of ordering of oxygen vacancies. The parent and the modulated spots have been indicated by arrows.

Fig. 5: SEM micrographs revealing the formation of ZnO islands obtained for 2 mins. ZnO deposition. Notice that the islands are separate and not connected.

Fig. 6: SEM micrographs exhibiting the ZnO nanostructure in the form of dense tetrapod networks as obtained from region 'A'. Notice the presence of several layers of tetrapod networks.

Fig. 7: A magnified view of the tetrapod networks. Some of the tetrapods for the two layers (upper and lower) have been marked by 'U' and 'L'.

Fig. 8: ZnO nanostructure obtained from region B (SEM). The growth of random nanorods around the periphery of the central region produces the flower like configuration.

Fig. 9: SEM micrographs depicting out the ordered nanorods in the regions, which are marked as NR and NR′. In the central region nanorods, which seem to be aligned/inclined along the electron-beam producing nearly end-on view can be seen. Some nanoparticles in the central region can also been seen.

**Table Caption.**

Lattice parameters of ZnO deposited at regions (A) and (B). Standard lattice parameters of ZnO from JCPDS file 36-1451 are a=3.24980Å and c=5.1948Å.

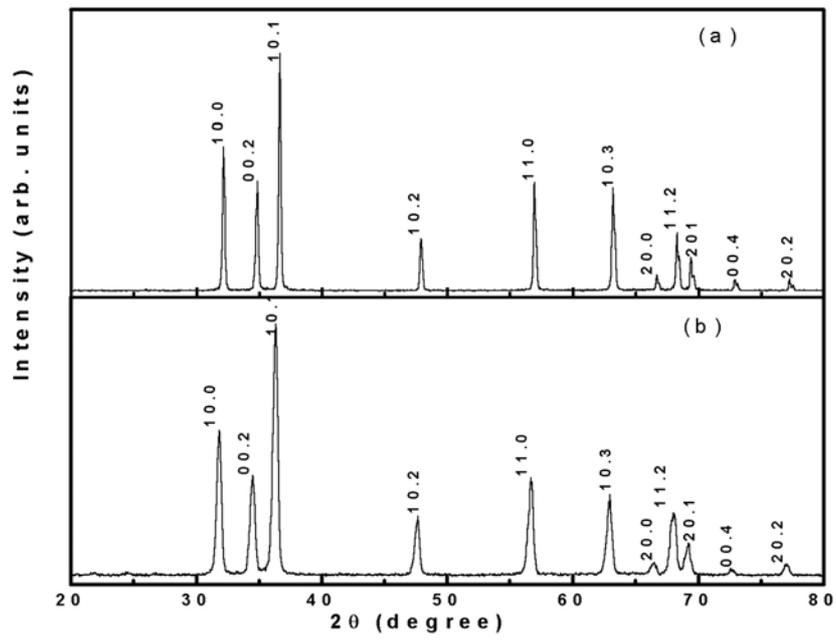

**Fig. 1**

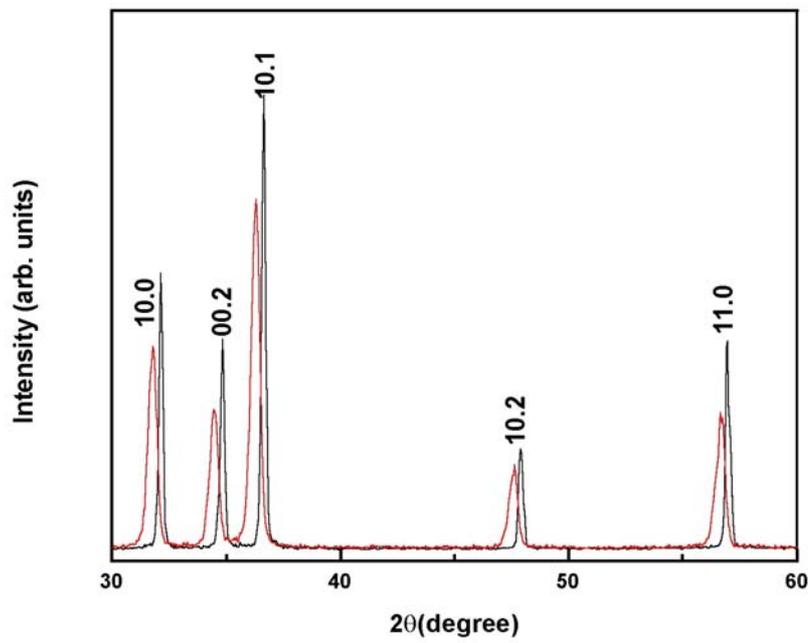

**Fig.2**

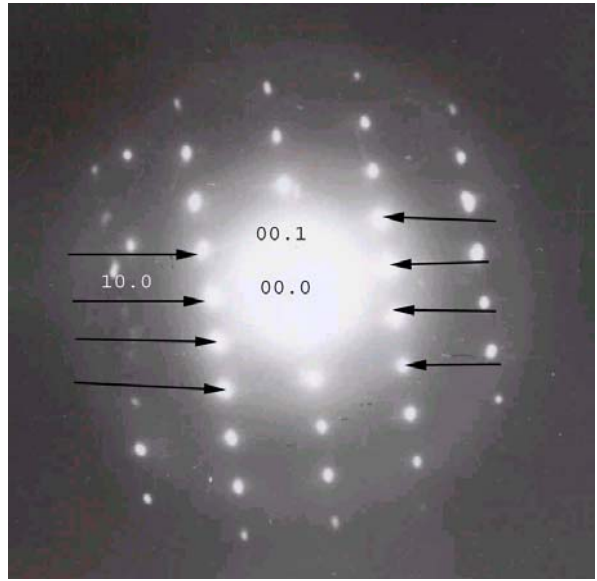

**Fig. 3**

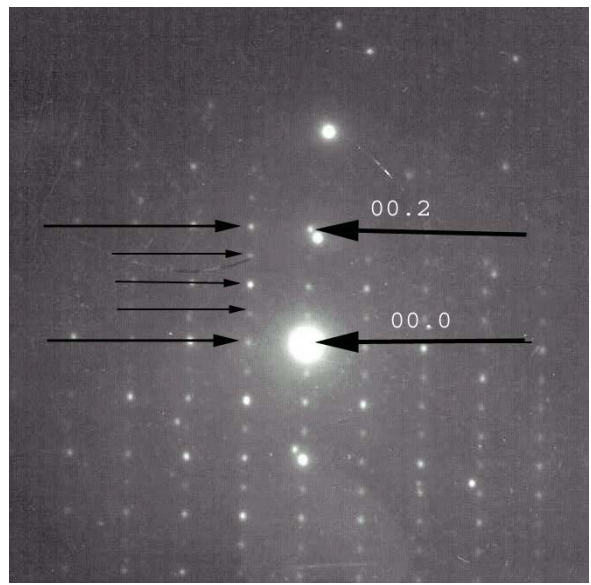

**Fig. 4**

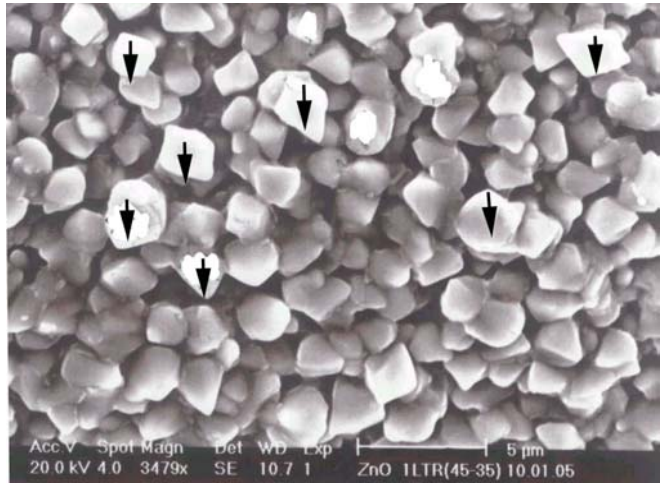

**Fig. 5**

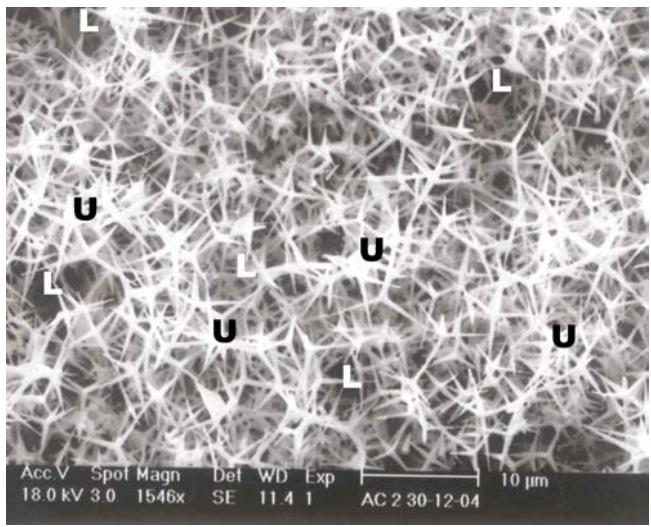

**Fig. 6**

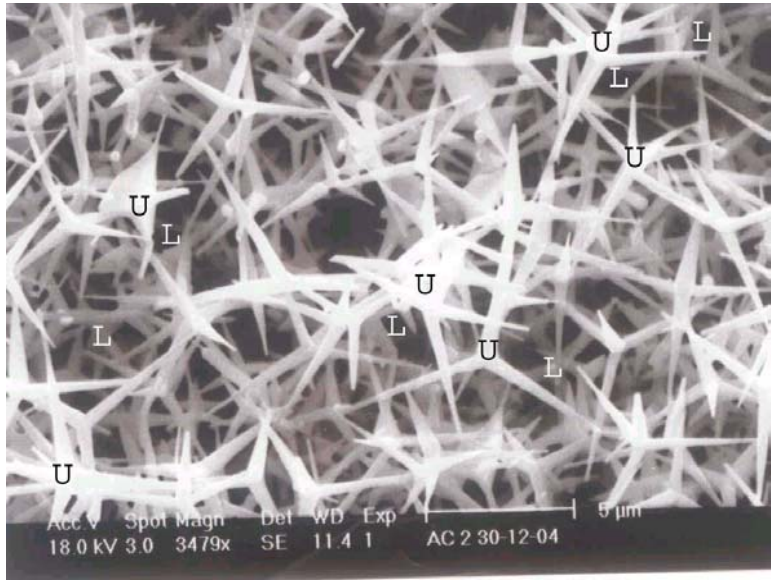

**Fig. 7**

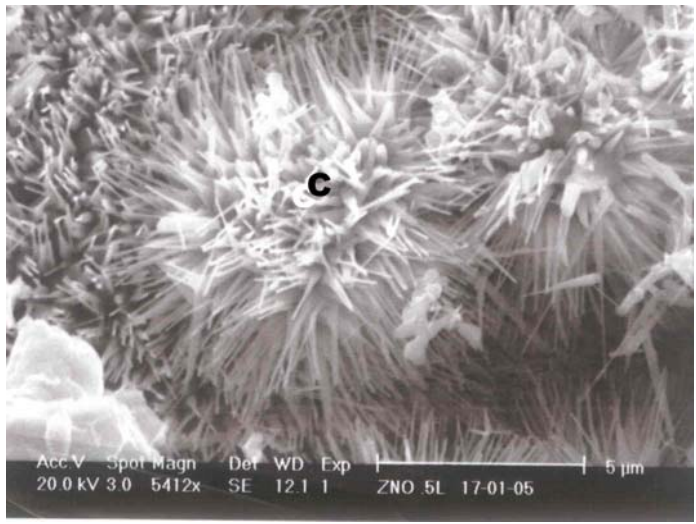

**Fig. 8**

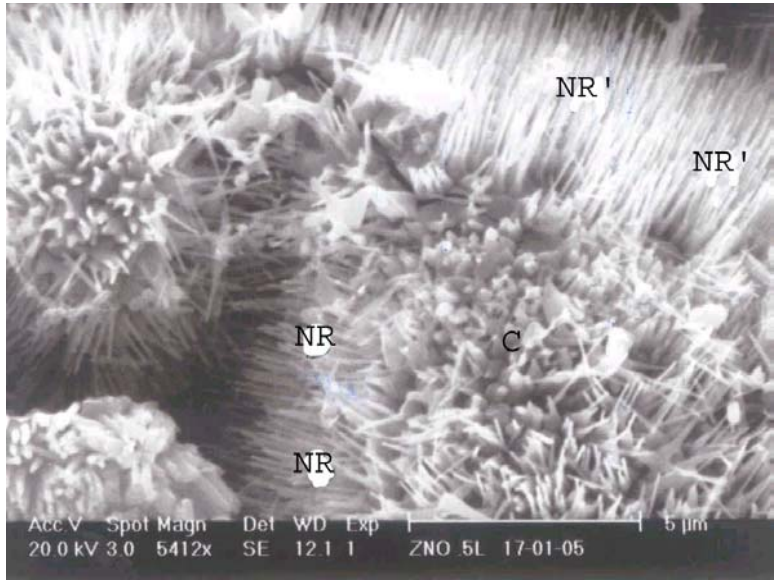

**Fig. 9**